# JOINT RATE ALLOCATION WITH BOTH LOOK-AHEAD AND FEEDBACK MODEL FOR HIGH EFFICIENCY VIDEO CODING

*Hongfei Fan, Lin Ding, Xiaodong Xie, Huizhu Jia and Wen Gao, Fellow, IEEE*

Institute of Digital Media, School of Electronic Engineering and Computer Science,
Peking University, Beijing100871, China
hffan@pku.edu.cn; ding.lin@pku.edu.cn; donxie@pku.edu.cn; hzjia@pku.edu.cn; wgao@pku.edu.cn

**ABSTRACT**

The objective of joint rate allocation among multiple coded video streams is to share the bandwidth to meet the demands of minimum average distortion (*minAVE*) or minimum distortion variance (*minVAR*). In previous works on *minVAR* problems, bits are directly assigned in proportion to their complexity measures and we call it look-ahead allocation model (LAM), which leads to the fact that the performance will totally depend on the accuracy of the complexity measures. This paper proposes a look-ahead and feedback allocation model (LFAM) for joint rate allocation for High Efficiency Video Coding (HEVC) platform which requires negligible computational cost. We derive the model from the target function of *minVAR* theoretically. The bits are assigned according to the complexity measures, the distortion and bitrate values fed back by the encoder together. We integrated the proposed allocation model in HEVC reference software HM16.0 and several complexity measures were applied to our allocation model. Results demonstrate that our proposed LFAM performs better than LAM, and an average of 65.94% variance of mean square error (MSE) is saved with different complexity measures.

## 1. INTRODUCTION

In most multimedia applications, different video sequences are encoded independently and transmitted over a bandwidth-limited channel. However, independent encoding may lead to a significant difference in quality among videos. Statistical multiplexing is a way to multiplex multiple video streams limited by an overall bandwidth, and the bandwidth is allocated due to the different characteristics of video streams. Typically, the quality of high complexity videos is improved at the expense of a reduction in the quality of low complexity videos. How to share the bandwidth among multiple sequences is a challenge. Basically, the existing methods can be classified into minimum average distortion (*minAVE*) and minimum distortion variance (*minVAR*).

The *minAVE* methods aim to achieve maximum average visual quality among multiple videos. [2]-[5] have proposed different methods for *minAVE* problems. While the *minVAR* methods such as [6]-[10] aim to achieve minimum variance in visual quality among multiple sequences, which means to achieve equal visual quality among all streams. Although *minAVE* methods will lead to a better overall visual quality, *minVAR* is still a more preferred solution in many applications such as broadcasting. In this paper, we focus on *minVAR* problems.

In order to achieve *minVAR*, many works [6]-[10] proposed different complexity measures and the bandwidth was assigned to each stream in proportion to their complexity measures and we call it look-ahead allocation model (LAM). In [6], average activity was applied to characterize scene content. In [7], the complexity measure was determined by the number of macroblocks, the sum of motion vector components, and the mean absolute difference (MAD) together. In [8], a novel complexity measure that adapts to the characteristics of H.264 video coding was proposed. In [9], both the frame activity and the motion activity were used to characterize the video complexity. In [10], a new complexity measure which incorporates the characteristics of Human Visual System (HVS) was introduced in the look-ahead approach. According to these works, a scheme performed better with a more precise complexity measure. However, all these methods assigned bits in proportion to complexity measures directly with no theoretical support and the allocation performance of LAM totally depends on the accuracy of the complexity measures.

In this paper, we propose a look-ahead and feedback allocation model (LFAM) for joint rate allocation. The proposed model is derived from the target function of the *minVAR* problem considering not only complexity measures but also the bitrate and distortion values fed back from the encoder. The group of frames (GOP) of all the sequences encoded in a fixed time interval is called a super GOP in this paper. Firstly, the bitrate and distortion values of the last coded super GOPs are fed back from the encoder. Then, the complexity measures of the next super GOPs are obtained by look-ahead approach. After that, bits are assigned to each sequence according to LFAM considering the feedback bitrate, distortion, and look-ahead complexity measure together. Finally, the super GOPs of different sequences are coded with allocated bitrate. We applied all the complexity measures in [6]-[10] to our allocation model and results show that our proposed LFAM completely outperforms the complexity directly assigned by LAM.

The rest of this paper is organized as follows. Section II gives the theoretical support of the proposed LFAM in detail. Section III shows the experimental results to demonstrate the effectiveness of LFAM and this paper is concluded in section V.

## 2. FORMATTING YOUR PAPER

### 2.1. Problem Formulation and Current Solution

All $N$ sequences are multiplexed into a single channel and we have the rate constraint for every super GOP as,

$$\sum_{i=0}^{N} R_k^i \leq R_c \qquad (1)$$

where $R_c$ denotes the total bandwidth for a super GOP of N videos, $R_k^i$ denotes the bitrate of the $i_{th}$ video in the $k_{th}$ super GOP, $N$ denotes the total number of videos. Since we focus on the *minVAR* problem, considering the rate constraint in (1), the question can be formulated as,

$$\min_{R_k^i} \left\{ \sum_{i=1}^{N} |D_k^i - \overline{D_k}| \right\} \quad s.t. \sum_{i=1}^{N} R_k^i \leq R_c \qquad (2)$$

where $D_k^i$ denotes the distortion of the $k_{th}$ super GOP of the $i_{th}$ video, and $\overline{D_k}$ denotes the average distortion of the $k_{th}$ super GOP of N videos. In this paper, the average mean square error (MSE) is used to measure the distortion.

LAM applied in related works [6]-[10] to solve *minVAR* problems can be summarized as,

$$R_{k+1}^i = \frac{C_{k+1}^i}{\sum_{i=1}^{N} C_{k+1}^i} \cdot R_c \qquad (3)$$

where $R_{k+1}^i$ denotes the bitrate of the $i_{th}$ video in the $(k+1)_{th}$ super GOP, $C_{k+1}^i$ denotes the complexity measure of the $i_{th}$ video in the $(k+1)_{th}$ super GOP.

The accuracy of LAM for *minVAR* problems depends on the accuracy of the complexity measures. Meanwhile, whether assigning bits in proportion to complexity can achieve equal visual quality among all streams or not is uncertain. To solve these problems, we derive the proposed allocation model from the target function of *minVAR* problem in (2). Details are given in the following.

### 2.2. Relationship between Rate and Distortion

In [11] and [12], the relationship among the Lagrange multiplier, the distortion and the bitrate was derived as,

$$\lambda = -\frac{\partial D}{\partial R} \qquad (4)$$

where $\lambda$ represents the Lagrange multiplier, $D$ denotes distortion and $R$ denotes bitrate.

The computation method of the Lagrange multiplier in HEVC encoder is defined as,

$$\lambda = cQ^2 \qquad (5)$$

where $Q$ is *QPstep* and $c$ is a constant determined by experimental results.

A well-known quadratic relationship between $R$ and $Q$ was modeled in [1] as,

$$R = \frac{a \cdot C}{Q} + \frac{b \cdot C^2}{Q^2} \qquad (6)$$

where $C$ is the complexity measure of the stream. $a$ and $b$ are constant values. Since (6) was derived by expanding rate distortion function into a Taylor series, a linear relationship can also be used for a coarse estimation of the relationship between $R$ and $Q$ as,

$$R = \frac{a \cdot C}{Q} \qquad (7)$$

Combine (5) and (7), a relationship between $\lambda$ and $R$ can be derived as,

$$\lambda = \frac{\sigma \cdot C^2}{R^2} \qquad (8)$$

where $\sigma$ is a constant value which changes according to the characteristics of the sequence. Since $\lambda$ is the slope of rate-distortion curve which is described in (4), so,

$$-\frac{\partial D}{\partial R} = \sigma \cdot \frac{C^2}{R^2} \qquad (9)$$

We can derive a relationship from (9) between $R$ and $D$ as,

$$D = \frac{\sigma \cdot C^2}{R} \qquad (10)$$

We verified (10) by experiments in HEVC reference software HM 16.0 with low delay configuration. QP value was set as 17, 22, 27, 32, 37, 42, and 47. $R$ was expressed in terms of bitrate and $D$ was expressed in terms of MSE of luma component. The result is given in Fig.1. In Fig.1, the horizontal axis and vertical axis represent bitrate and 1/MSE respectively. Test sequences recommended by JCT-VC were tested and the four results of them are given in Fig.1. Results show that the inverse relationship between $D$ and $R$ in (10) is acceptable.

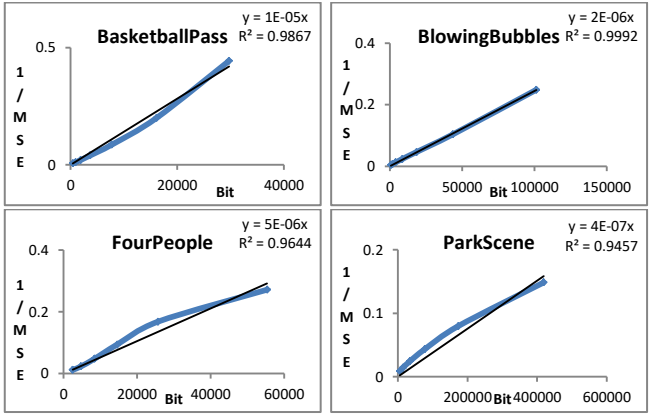

**Fig. 1.** Relationship between Rate and Distortion

### 2.3. Proposed Allocation Model

To derive the allocation model, we assume that the $k_{th}$ super GOP is coded, the complexity measures of the $k_{th}$ and $(k+1)_{th}$ super GOP of the $i_{th}$ video denoted as $C_k^i$ and $C_{k+1}^i$ are calculated with look-ahead approaches. The bitrate $R_k^i$ and distortion $D_k^i$ of the super GOP are obtained from the encoder. In this part, distortion is always measured by MSE. Considering the target function given in (1), the best allocation for the $(k+1)_{th}$ super GOP is as follows,

$$\forall_{i,j}\ D_{k+1}^i = D_{k+1}^j \quad s.t. \sum_{i=1}^{N} R_{k+1}^i = R_c \quad (11)$$

According to the relationship given in (10), we have,

$$D_k^i = \frac{\sigma_k^i \cdot {C_k^i}^2}{R_k^i} \quad (12)$$

Since $\sigma_k^i$ is a constant value which changes according to the characteristics of the sequence, we assume that the characteristics of a sequence are similar between two adjacent super GOPs. So,

$$\widehat{\sigma_{k+1}^i} \approx \sigma_k^i = \frac{D_k^i \cdot R_k^i}{{C_k^i}^2} \quad (13)$$

where $\widehat{\sigma_{k+1}^i}$ is the estimate value of $\sigma_{k+1}^i$. Combine (10) and (13), we can obtain,

$$\widehat{R_{k+1}^i} = \frac{\widehat{\sigma_{k+1}^i} \cdot {C_{k+1}^i}^2}{\widehat{D_{k+1}^i}} \quad (14)$$

where $\widehat{D_{k+1}^i}$ is the estimate value of $D_{k+1}^i$ and $\widehat{R_{k+1}^i}$ denotes the estimation of the allocation bits.

Considering $\forall_{i,j}\ D_{k+1}^i = D_{k+1}^j$, we sum up $\widehat{R_{k+1}^i}$ together,

$$R_c \approx \sum_{i=1}^{N} \widehat{R_{k+1}^i} = \frac{\sum_{i=1}^{N}(\widehat{\sigma_{k+1}^i} \cdot {C_{k+1}^i}^2)}{\widehat{D_{k+1}^1}} \quad (15)$$

We divide (14) by (15),

$$\frac{\widehat{R_{k+1}^i}}{R_c} = \frac{\widehat{\sigma_{k+1}^i} \cdot {C_{k+1}^i}^2}{\sum_{i=1}^{N}(\widehat{\sigma_{k+1}^i} \cdot {C_{k+1}^i}^2)} \quad (16)$$

Combine (13) and (16), we finally allocate $R_{k+1}^i$ as,

$$\widehat{R_{k+1}^i} = \frac{X_{k+1}^i}{\sum_{i=1}^{N} X_{k+1}^i} \cdot R_c$$

where,

$$X_{k+1}^i = \frac{D_k^i \cdot R_k^i \cdot {C_{k+1}^i}^2}{{C_k^i}^2} \quad (17)$$

The model in (17) is a look-ahead and feedback allocation model (LFAM). The calculation of C has been given in many works [6]-[10]. All the complexity measures in the related works can fit LFAM. We verified our model with different complexity measures in the next section.

### 3. EXPERIMENTAL RESULTS

We integrated our proposed LFAM in HEVC reference software HM 16.0. Low delay configuration was used. Both frame level and LCU level rate control were turned on. The frame rate was set as 25 fps. Test sequences recommended by JCT-VC were tested and were divided into 6 sets according to different classes. The total bitrate $R_c$ were set as 20M, 10M, 4M, 1M, 2M, and 6M per second for Class A to Class F, respectively. The sequence number is 2, 5, 4, 4, 3, and 3 for Class A to Class F, respectively. 300 frames were tested for Class C, Class D, Class E, and Class F. 240 frames were tested for Class B and 150 frames for Class A. The length of a super GOP was set as 10 in the following experiments. Variance is a measure of how far a set of numbers is spread out in statistics. Thus a smaller variance of MSE represents a better efficiency of the allocation model, and the variance is defined as,

$$Variance_k = \sum_{i=1}^{N}\left(MSE_k^i - \overline{MSE_k}\right)^2 \quad (18)$$

where $Variance_k$ represents the variance of MSE of all sequences in the $k_{th}$ super GOP. $MSE_k^i$ represents the MSE

**Table 1.** Comparison of variances between LAM and LFAM

| Complexity Measure | Allocation Model | Stream Sets | | | | | | Average |
|---|---|---|---|---|---|---|---|---|
| | | Class A | Class B | Class C | Class D | Class E | Class F | |
| C[6] | LAM | 31.01 | 56.57 | 462.31 | 249.83 | 0.76 | 30.02 | 138.42 |
| | LFAM | 4.63 | 7.82 | 65.45 | 94.21 | 0.4 | 10.65 | 30.53 |
| | Saving | 85.08% | 86.18% | 85.84% | 62.29% | 47.37% | 64.52% | **71.88%** |
| C[7] | LAM | 18.64 | 27.44 | 377.65 | 199.87 | 0.47 | 41.96 | 111.01 |
| | LFAM | 1.2 | 6.18 | 58.86 | 69.65 | 0.35 | 5.72 | 23.66 |
| | Saving | 93.54% | 77.48% | 84.41% | 65.15% | 25.53% | 86.37% | **72.08%** |
| C[8] | LAM | 21.69 | 39.66 | 407.03 | 510.36 | 0.35 | 39.03 | 169.69 |
| | LFAM | 4.7 | 6.36 | 40.46 | 62.11 | 0.25 | 10.15 | 20.67 |
| | Saving | 78.31% | 83.96% | 90.06% | 87.83% | 28.57% | 73.99% | **73.79%** |
| C[9] | LAM | 26.67 | 16.7 | 286.56 | 254.73 | 0.45 | 26.8 | 101.98 |
| | LFAM | 11.19 | 9.63 | 60.38 | 102.79 | 0.35 | 11.14 | 32.58 |
| | Saving | 58.03% | 42.34% | 78.93% | 59.65% | 22.22% | 58.43% | **53.27%** |
| C[10] | LAM | 18.12 | 60.2 | 277.89 | 137.31 | 0.54 | 9.8 | 83.98 |
| | LFAM | 4.44 | 9.55 | 48.73 | 99.79 | 0.35 | 5.15 | 28 |
| | Saving | 75.49% | 84.14% | 82.46% | 27.33% | 35.19% | 47.45% | **58.68%** |
| Average | LAM | 23.23 | 40.11 | 362.29 | 270.42 | 0.51 | 29.52 | 121.01 |
| | LFAM | 5.23 | 7.91 | 54.78 | 85.71 | 0.34 | 8.56 | 27.09 |
| | Saving | 77.47% | 80.29% | 84.88% | 68.31% | 33.85% | 71.00% | **65.94%** |

**Table 2.** Average PSNR and Variance for Each Super GOP at Class C applying *C[10]* as complexity measure

| GOP Index | Average PSNR | | | | | | | | Variance | |
|---|---|---|---|---|---|---|---|---|---|---|
| | LAM | | | | LFAM | | | | LAM | LFAM |
| | Basketball Drill | BQMall | PartyScene | RaceHorses | Basketball Drill | BQMall | PartyScene | RaceHorses | | |
| 1 | 37.31 | 35.67 | 30.79 | 31.41 | 36.09 | 32.92 | 33.11 | 31.46 | 330.65 | 116.82 |
| 2 | 36.82 | 35.34 | 30.47 | 31.16 | 34.67 | 31.68 | 33 | 31.29 | 370.58 | 104.51 |
| 3 | 36.35 | 35.09 | 30.22 | 31.08 | 33.73 | 31.19 | 32.36 | 31.33 | 392.21 | 77.61 |
| 4 | 36.04 | 35.03 | 29.96 | 31.13 | 33.37 | 31.41 | 31.92 | 32.04 | 423.08 | 38.65 |
| 5 | 35.96 | 35.08 | 29.66 | 30.97 | 33.5 | 31.93 | 31.56 | 31.96 | 503.26 | 38.01 |
| 6 | 36.26 | 35.18 | 30.13 | 30.97 | 33.84 | 32.65 | 31.62 | 31.98 | 417.03 | 45.9 |
| 7 | 36.73 | 35.31 | 30.87 | 31.74 | 33.98 | 33.31 | 32 | 31.9 | 269.69 | 47.03 |
| 8 | 36.83 | 35.76 | 31.82 | 31.76 | 33.75 | 33.84 | 32.42 | 31.61 | 193.6 | 55.67 |
| 9 | 36.8 | 35.86 | 31.87 | 30.57 | 33.52 | 33.87 | 32.37 | 30.67 | 325.62 | 130.81 |
| 10 | 36.27 | 35.67 | 31.57 | 29.11 | 33.19 | 33.69 | 31.84 | 29.85 | 678.49 | 239.54 |
| 11 | 35.73 | 35.09 | 30.88 | 29.52 | 32.94 | 33.04 | 31.6 | 30.59 | 534.34 | 100.48 |
| 12 | 35.32 | 34.98 | 30.64 | 30.96 | 32.83 | 33.11 | 32.21 | 32.4 | 295.46 | 8.16 |
| Avg | **36.37** | **35.34** | **30.74** | **30.87** | **33.78** | **32.72** | **32.17** | **31.42** | 364.15 | 77.17 |

**Table 3.** Average PSNR and Variance for Each Super GOP at Class B applying *C[10]* as complexity measure

| Gop Index | Average PSNR | | | | | | | | | | Variance | |
|---|---|---|---|---|---|---|---|---|---|---|---|---|
| | LAM | | | | | LFAM | | | | | LAM | LFAM |
| | Basketball Drive | BQTerrace | Cactus | Kimono | ParkScene | Basketball Drive | BQTerrace | Cactus | Kimono | ParkScene | | |
| 1 | 34.39 | 35.36 | 33.87 | 36.78 | 33.88 | 35.71 | 34.49 | 34.65 | 36.12 | 34.47 | 25.09 | 9.55 |
| 2 | 34.10 | 34.31 | 33.81 | 36.87 | 32.99 | 35.14 | 33.74 | 34.76 | 35.81 | 34.29 | 39.57 | 12.80 |
| 3 | 34.17 | 33.24 | 33.94 | 36.82 | 32.39 | 34.28 | 33.32 | 34.71 | 35.48 | 34.49 | 61.91 | 14.96 |
| 4 | 34.28 | 32.76 | 34.02 | 36.79 | 32.08 | 34.30 | 33.40 | 34.69 | 35.22 | 34.45 | 83.52 | 11.31 |
| 5 | 34.32 | 32.47 | 34.02 | 36.85 | 31.82 | 34.09 | 33.92 | 34.57 | 34.92 | 34.68 | 105.38 | 4.15 |
| 6 | 34.15 | 32.91 | 33.89 | 36.93 | 31.88 | 34.36 | 34.65 | 34.57 | 34.75 | 34.49 | 92.19 | 0.50 |
| 7 | 34.32 | 33.47 | 33.95 | 37.06 | 32.10 | 34.33 | 35.17 | 34.44 | 34.57 | 34.65 | 77.17 | 2.12 |
| 8 | 34.82 | 34.25 | 34.12 | 37.21 | 32.66 | 34.83 | 35.44 | 34.67 | 34.65 | 34.79 | 54.01 | 1.86 |
| 9 | 35.00 | 34.46 | 34.40 | 37.47 | 33.16 | 35.70 | 35.72 | 34.89 | 34.89 | 35.34 | 40.33 | 2.66 |
| 10 | 35.12 | 34.66 | 34.59 | 37.59 | 33.42 | 36.53 | 35.96 | 35.39 | 35.29 | 35.69 | 34.42 | 2.93 |
| 11 | 34.40 | 34.40 | 34.70 | 37.56 | 33.68 | 35.88 | 35.77 | 35.54 | 35.54 | 36.23 | 30.30 | 1.01 |
| 12 | 34.91 | 33.57 | 34.71 | 37.46 | 33.69 | 35.78 | 34.51 | 35.48 | 35.92 | 36.02 | 36.92 | 6.12 |
| Avg | **34.50** | **33.82** | **34.17** | **37.12** | **32.81** | **35.08** | **34.67** | **34.86** | **35.26** | **34.97** | 52.37 | 5.38 |

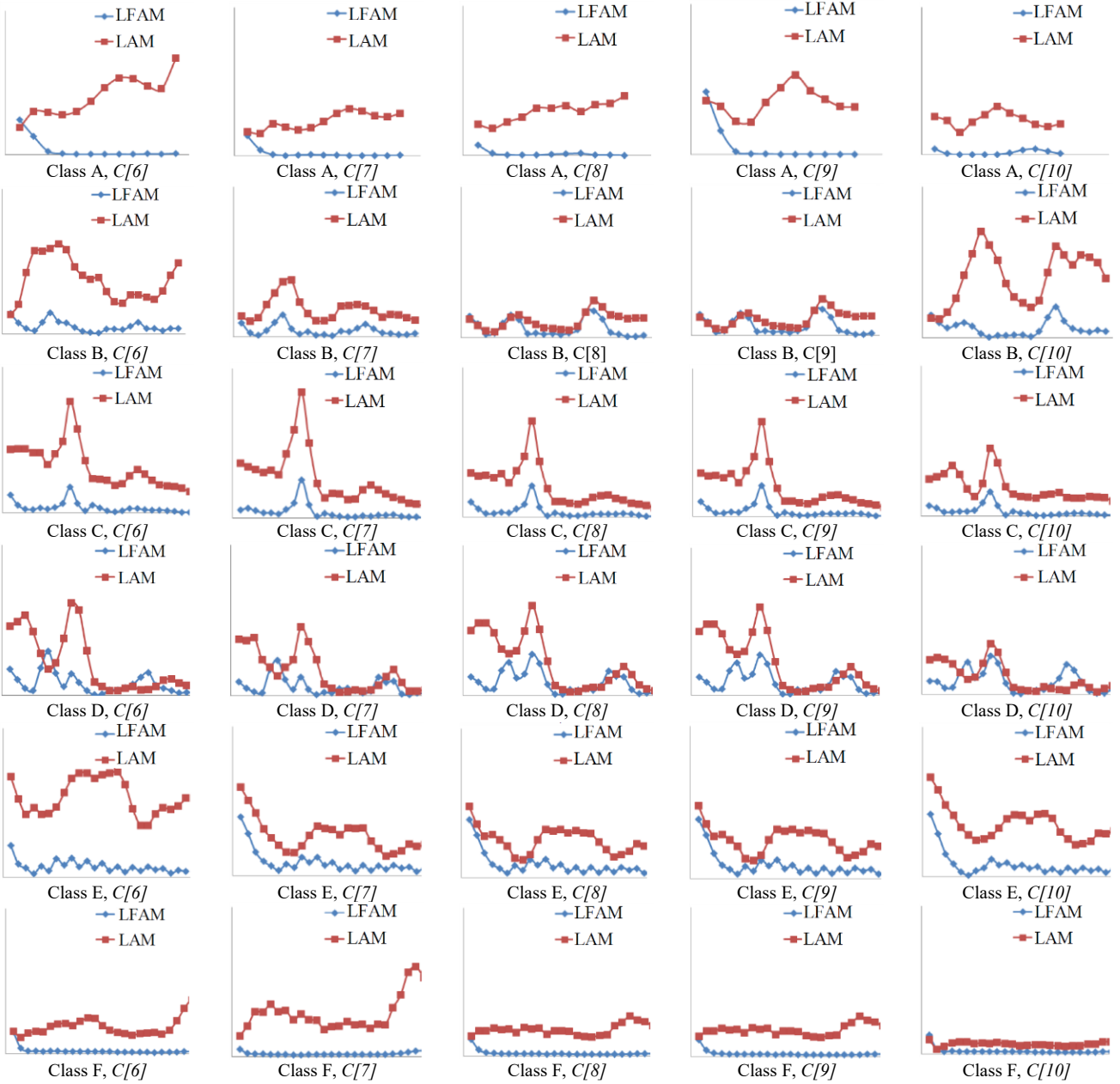

**Fig. 2.** Comparison of variance value between proposed LFAM and related LAM with different complexity measures at different classes

value of the $i_{th}$ video in the $k_{th}$ super GOP and $\overline{MSE_k}$ is the average value of different videos in the $k_{th}$ super GOP.

The complexity measures derived in [6]-[10] were applied in our experiment which are denoted as $C[6]$-$C[10]$. Since not all the constant values of $C[6]$-$C[10]$ were given in the corresponding works, these values were set based on the author's understanding. So, the comparison between different complexity measures in our experiment was not totally fair because the efficiency of these measures is affected by the mentioned values. However, the comparison between the LAM and the proposed LFAM was fair because same values were applied in our experiments.

Table 1 shows the comparison of the average MSE variance between LAM and LFAM applying different complexity measures. Bitrates of the first super GOPs for different sequence were initialized as the same and variances of the first super GOPs were removed in the following results. In the 'Complexity Measure' column, $C[6]$-$C[10]$ represent the complexity measures given in [6]-[10] and 'Average' denotes the average variance value of different complexity measures. In 'Allocation Model' column, 'LAM' denotes the model given in (3) and 'LFAM' denotes the model given in (17), and 'Saving' denotes the variance saved by applying LFAM and was defined as,

$$Saving = \frac{Variance_{LAM} - Variance_{LFAM}}{Variance_{LAM}} \quad (19)$$

where $Variance_{LAM}$ denotes the variance of MSE applying LAM and $Variance_{LFAM}$ denotes that applying LFAM.

The 'Stream Sets' column denotes the test sequence sets. The 'Average' column gives the average variance value of the different stream sets. As a result, the overall average 'Saving' is 65.94%. The proposed LFAM performs much better with different complexity measures for all classes which means the accuracy of the model is not fully depends on the accuracy of the complexity measure.

For a more accurate comparison, Table 2 and Table 3 give the average PSNR and variance for each super GOP at Class C and Class B. The complexity measure given in [10] was applied. The result of first GOP was not shown in the table and results of first 12 super GOPs are given. In the test set of Class C, the average PSNRs of sequence RaceHorses and PartyScene were 30.87 and 30.74, while that of the other two sequences were 36.37 and 35.34. The PSNR of sequence RaceHorses and PartyScene were lower than the other two sequences when applying the related model. With the proposed allocation model, the PSNR of RaceHorses and PartyScene were 31.42 and 32.17, while that of the other two sequences were 33.78 and 32.72. The PSNR of RaceHorses and PartyScene were improved at the expense of a reduction in the quality of the other two videos. In the test set of Class B, the average PSNR of Kimono was 37.12 and the PSNR of ParkScene was 32.81 with related model. The quality between Kimono and ParkScene was quite large. While applying the proposed model, the average PSNR of Kimono and ParkScene were 35.26 and 34.97, respectively. The quality between Kimono and ParkScene was almost same when applying the proposed model. As a result, the variance is stable when applying the proposed model.

Fig.2 shows the performance of LFAM and LAM by diagrams. The horizontal axis and vertical axis denote super GOP index and variance value, respectively. All the results with different complexity measures at different classes are given. Each point in the diagram represents the variance result of one super GOP. As given in the diagrams, the variances between two models were similar at the beginning. After a few super GOPs, the variance applying LFAM rapidly decreased and remained stable at a low level. Our proposed LFAM performs much better than LAM under all the situations.

Since the related model assigns bits in proportion to complexity measures directly, the performance will depend on the accuracy of the complexity measure. Considering the mentioned results, the complexity measures cannot totally represent the coding complexity. Meanwhile, the improvement of the accuracy of the complexity measure is mostly at the expense of a higher computational cost. Thus, the proposed look-ahead and feedback allocation model is a better solution to supplement the inaccuracy of the complexity measure with only little extra computational cost.

## 4. CONCLUSION

A look-ahead and feedback allocation model denoted as LFAM for statistic multiplexing is proposed in this paper. Experimental results based on HEVC platform show that the accuracy of statistic multiplexing does not fully depend on the accuracy of the complexity measure anymore. Considering the complexity measures, the distortion and bitrate values fed back by the encoder, LFAM performs much better than LAM applied in the related works. The theoretical support is given and only little computational cost is required. Totally, 65.94% variance of MSE is saved in average applying the proposed LFAM.